\documentclass[twocolumn,prl,superscriptaddress]{revtex4}
\usepackage{amsmath,amssymb,amsfonts,mathrsfs}
\usepackage{bm}
\usepackage{graphicx}
\usepackage{hyperref}
\usepackage{amsbsy}
\usepackage{dcolumn}

\hypersetup{
     colorlinks = true,
     citecolor  = red,  
     linkcolor  = blue  
}

\begin{document}

\title{Experimental Observation of Tensor Monopoles with a Superconducting Qudit}

\author{Xinsheng Tan}
\email{tanxs@nju.edu.cn}
\affiliation{National Laboratory of Solid State Microstructures, School of Physics,
Nanjing University, Nanjing 210093, China}
\author{Dan-Wei Zhang}
\email{danweizhang@m.scnu.edu.cn}
\affiliation{Guangdong Provincial Key Laboratory of Quantum Engineering and Quantum Materials, School of Physics and Telecommunication Engineering, South China Normal University, Guangzhou 510006, China}
\affiliation{Guangdong-Hong Kong Joint Laboratory of Quantum Matter, Frontier Research Institute for Physics, South China Normal University, Guangzhou 510006,
China}


\author{Wen Zheng}
\affiliation{National Laboratory of Solid State Microstructures, School of Physics,
Nanjing University, Nanjing 210093, China}
\author{Xiaopei Yang}
\affiliation{National Laboratory of Solid State Microstructures, School of Physics,
Nanjing University, Nanjing 210093, China}
\author{Shuqing Song}
\affiliation{National Laboratory of Solid State Microstructures, School of Physics,
Nanjing University, Nanjing 210093, China}
\author{Zhikun Han}
\affiliation{National Laboratory of Solid State Microstructures, School of Physics,
Nanjing University, Nanjing 210093, China}
\author{Yuqian Dong}
\affiliation{National Laboratory of Solid State Microstructures, School of Physics,
Nanjing University, Nanjing 210093, China}
\author{Zhimin Wang}
\affiliation{National Laboratory of Solid State Microstructures, School of Physics,
	Nanjing University, Nanjing 210093, China}

\author{Dong Lan}
\affiliation{National Laboratory of Solid State Microstructures, School of Physics,
Nanjing University, Nanjing 210093, China}

\author{Hui Yan}
\affiliation{Guangdong Provincial Key Laboratory of Quantum Engineering and Quantum Materials, School of Physics and Telecommunication Engineering, South China Normal University, Guangzhou 510006, China}
\affiliation{Guangdong-Hong Kong Joint Laboratory of Quantum Matter, Frontier Research Institute for Physics, South China Normal University, Guangzhou 510006,
China}

\author{Shi-Liang Zhu}
\email{slzhu@nju.edu.cn}
\affiliation{Guangdong Provincial Key Laboratory of Quantum Engineering and Quantum Materials, School of Physics and Telecommunication Engineering, South China Normal University, Guangzhou 510006, China}
\affiliation{Guangdong-Hong Kong Joint Laboratory of Quantum Matter, Frontier Research Institute for Physics, South China Normal University, Guangzhou 510006,
China}

\author{Yang Yu}
\email{yuyang@nju.edu.cn}
\affiliation{National Laboratory of Solid State Microstructures, School of Physics,
Nanjing University, Nanjing 210093, China}

\begin{abstract}
Monopoles play a center role in gauge theories and topological matter. There are two fundamental types of monopoles in physics: vector monopoles and tensor monopoles. Examples of vector monopoles include the Dirac monopole in 3D and Yang monopole in 5D, which have been extensively studied and observed in condensed matter or artificial systems. However, tensor monopoles are less studied, and their observation has not been reported. Here we experimentally construct a tunable spin-1 Hamiltonian to generate a tensor monopole and then measure its unique features with superconducting quantum circuits. The energy structure of a 4D Weyl-like Hamiltonian with three-fold degenerate points acting as tensor monopoles is imaged. Through quantum-metric measurements, we report the first experiment that measures the Dixmier-Douady invariant, the topological charge of the tensor monopole. Moreover, we observe topological phase transitions characterized by the topological Dixmier-Douady invariant, rather than the Chern numbers as used for conventional monopoles in odd-dimensional spaces.
\end{abstract}

\maketitle

\bigskip

\emph{Introduction.}--Monopoles are fundamental topological objects in high-energy physics and condensed matter physics. In 1931, Dirac captured the physical importance of magnetic monopoles (called Dirac monopoles) \cite{Dirac}, and proved the quantization of the electric charge. The Dirac monopole was later recognized to be connected to the Berry curvature and Berry phase in quantum mechanics \cite{Xiao}. The topological nature of Dirac monopoles defined in three dimensions (3D) is characterized by the first Chern number. Other monopoles have been identified in gauge theory, such as the 't Hooft-Polyakov monopole \cite{Polyakov,Hooft} in Yang-Mills theory and the Yang monopole \cite{Yang}. The Yang monopole is a non-Abelian extension of the Dirac monopole in five dimensions (5D) and is characterized by the second Chern number. Generally, a zoo of monopoles in ($2n+1$)-dimensional ($n=1,2,3...$) flat spaces can be identified by the $n$-order Chern numbers, which are given by the integral of the corresponding field strength associated with a monopole's gauge field \cite{Nakahara}.

From the aspect of gauge fields, there are two fundamental types of monopoles in physics: vector monopoles associated with vector gauge fields, such as the aforementioned Dirac and Yang monopoles, and tensor monopoles associated with tensor gauge fields ~\cite{Nepomechie,Teitelboim,Orland,Kalb}.
A representative of the so-called ``tensor monopole" is defined in a four-dimensional (4D) space. The topological charge of a 4D tensor monopole is given by the integral of the tensor gauge field ~\cite{Kalb,Palumbo2018,Palumbo2019,YQZhu2020}, known as the Dixmier-Douady (DD) invariant \cite{Mathai,Murray}. 
Tensor monopoles play a key role in string theory, where currents naturally couple to a tensor gauge field \cite{Banks2011,Mavromatos2017,Montero2017}.
Recently, Palumbo and Goldman proposed a realistic three-band model defined over a 4D parameter space to generate tensor monopoles \cite{Palumbo2018,Palumbo2019}, whose topological charges could be extracted from the generalized Berry curvature by measuring the quantum metric \cite{Kolodrubetz,Lim2015,Provost,Ma2010,Rezakhani}. 
The quantum metric in engineered quantum systems can be measured through periodic driving \cite{Ozawa2018,Cai}, sudden quench \cite{Tan2019b}, and spin-texture \cite{Bleu,Gianfrate}.

So far, monopoles have not been observed for real particles. However, they can emerge in condensed-matter materials \cite{Qi,Armitage} or be engineered in certain artificial systems with effective gauge fields \cite{Dalibard,Goldman,DWZhang2018,Ozawa}. In these systems, monopoles are usually connected to the existence of topological states. For instance, Weyl points in Weyl semimetals can be viewed as fictitious Dirac monopoles in momentum space \cite{Armitage}. The analog Dirac monopoles were created in the synthetic electromagnetic field that arises in the spin texture of atomic spinor condensates~\cite{Ray2014,Ray2015}. The monopole field and the first Chern number were measured in a 3D parameter space of spin-1/2 or spin-1 artificial atoms \cite{Schroer2014,Roushan2014,Tan2018,Tan2019a}. A quantum-simulated Yang monopole was observed in a 5D parameter space built from an atomic condensate's internal states, and the second Chern number as its topological charge was measured \cite{Sugawa}. Although the fundamental importance of singularity points associated with tensor gauge fields was theoretically revealed in high-energy physics and condensed matter physics \cite{Banks2011,Mavromatos2017,Montero2017,Mathai,Murray,Kalb,Palumbo2018,Palumbo2019,YQZhu2020}, the tensor monopoles have not yet been realized or simulated, and the corresponding topological DD invariant has not been measured.

\begin{figure}[tbph]
\includegraphics[width=7.5cm]{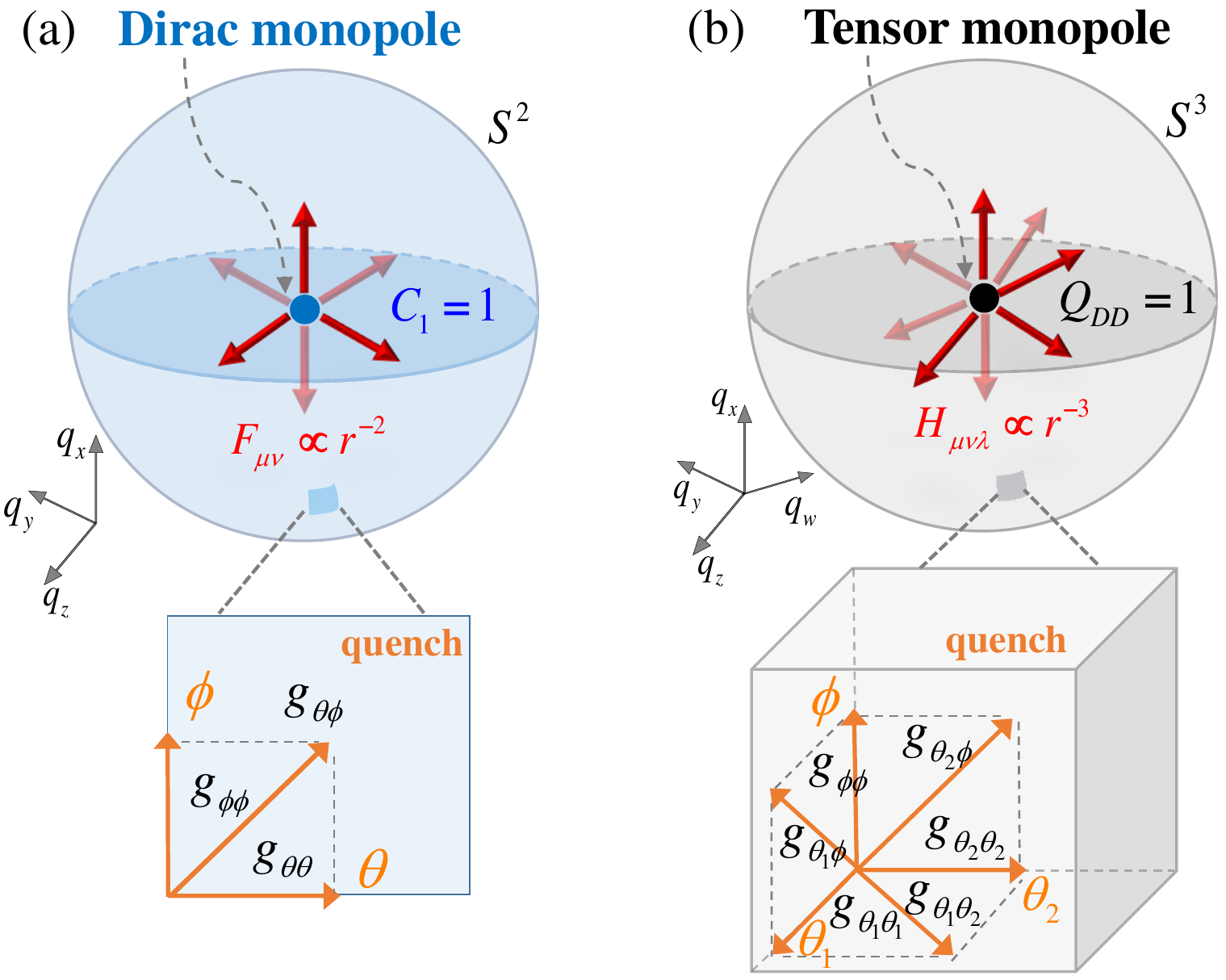}
\caption{(Color online) Pictorial representations of (a) a Dirac monopole in 3D parameter space $\boldsymbol q=(q_x,q_y,q_z)$; and (b) a tensor monopole in 4D parameter space $\boldsymbol q=(q_x,q_y,q_z,q_w)$. The two are defined as pointlike sources of vector and tensor gauge fields, respectively. The fluxes associated with the field strengths $\mathcal{F}_{\mu\nu}\propto r^{-2}$ and $\mathcal{H}_{\mu\nu\lambda}\propto r^{-3}$ through the surrounding 2D and 3D spheres ($S^2$ and $S^3$) with radius $r=|\mathbf{q}|$ are quantized in terms of two different topological invariants, the first Chern number $C_1=1$ and the DD invariant $Q_{DD}=1$, respectively. The related quantum metric tensors $g_{\mu\nu}$ in $S^2$ and $S^3$ can be measured from the quench scheme.}
\label{fig1}
\end{figure}

In this Letter, we fill this gap by experimentally synthesizing tensor monopoles in a 4D parameter space built in superconducting quantum circuits and measuring its topological features. By engineering a tunable 4D Weyl-like spin-1 Hamiltonian, we first image the energy structure with three-fold degenerate points acting as tensor monopoles. By characterizing the generalized curvature tensor through quantum-metric measurements, we report the first experiment to realize tensor gauge fields and measure the DD invariant as the topological charge of a tensor monopole. Finally, we engineer and observe the topological phase transition characterized by the DD invariant, where the manifold topology changes from a trivial state to a nontrivial one with the modification of a parameter in the Hamiltonian. Our work not only demonstrates the first observation of tensor monopoles and measurement of the DD invariant in a superconducting qudit, but also paves the way to explore high-dimensional topological defects in fully engineered quantum systems. The experimental observation of tensor monopoles can further our understanding of tensor gauge fields and advance the search for new exotic topological matter in condensed matter physics and artificial quantum systems.

\emph{Tensor monopoles and tensor fields.}--To establish a basic understanding of the tensor monopole in 4D parameter space, we begin by comparing it with the well-known Dirac monopole in 3D, both spanned by the parameters $\boldsymbol q$, as shown in Fig. \ref{fig1}. For a non-degenerate quantum state $|u_{{\boldsymbol q}}\rangle$, the geometric property is captured by a quantum geometric tensor \cite{Kolodrubetz,Carollo2020,SLZhu2008}: $\chi_{\mu\nu}=\langle \partial_{q_{\mu}} u_{\boldsymbol q} |(1 - | u_{\boldsymbol q}\rangle \langle u_{\boldsymbol q}|)| \partial_{q_{\nu}}u_{\boldsymbol q} \rangle=g_{\mu\nu}+i\mathcal{F}_{\mu\nu}/2$, where the real and imaginary parts define the quantum metric $g_{\mu\nu}=g_{\nu\mu}$ and Berry curvature (gauge field) $\mathcal{F}_{\mu\nu}=-\mathcal{F}_{\nu\mu}$, respectively. The Berry curvature $\mathcal{F}_{\mu\nu}=\!\partial_{\mu}A_{\nu}-\partial_{\nu}A_{\mu}$ with the Berry connection $A_{\mu}\!=i\!\langle u_{\boldsymbol q} \vert \partial_{q_{\mu}} u_{\boldsymbol q}\rangle$ is associated with the Berry phase. The quantum metric $g_{\mu\nu}$ defines the quantum distance between nearby states $|u_{\boldsymbol q}\rangle$ and $|u_{\boldsymbol q+d\boldsymbol q}\rangle$ in the parameter space \cite{Kolodrubetz,Lim2015,Provost,Ma2010,Rezakhani}: $ds^2=1-|\langle u_{\boldsymbol q}| u_{\boldsymbol q+\delta\boldsymbol q}\rangle|^2=\sum\nolimits_{\mu\nu} g_{\mu\nu} dq_{\mu} dq_{\nu}$,
which is related to the wave-function overlap and can thus be directly measured.

For a Dirac monopole in 3D $\boldsymbol q$ space, in the context of gauge field (electromagnetism), the Berry curvature $\mathcal{F}_{\mu\nu}$ can be viewed as the field strength (the Faraday tensor) associated with the flux through the surrounding sphere $S^2$ with  radius $r=|\boldsymbol q|$. A minimal model realizing a Dirac monopole is the Weyl Hamiltonian $H_{3D}=\boldsymbol q\cdot\boldsymbol{\sigma}$, where $\mathbf{\sigma}=(\sigma_x,\sigma_y,\sigma_z)$ are the Pauli matrices. The topological charge of the Dirac monopole at $\boldsymbol q=0$ is then given by the first Chern number $C_1=\frac{1}{2\pi}\int_{S^2} \mathcal{F}=1$. Notably, the Berry curvature associated with a monopole is related to the determinant of the metric tensor $g_{\mu\nu}$ defined on a sphere with $\mu,\nu=\{\theta,\phi\}$: $\mathcal{F}_{\mu\nu}= 2\epsilon_{\mu\nu}\sqrt{\det(g_{\mu\nu})}$, where $\epsilon_{\mu\nu}$ is the Levi-Civita symbol, $g_{\theta \theta} = 1/4$, $g_{\phi \phi} = \sin^2 \theta/4$, and $g_{\theta \phi}=0$.

Different from the odd-dimensional monopoles defined with vector fields, a tensor monopole is defined in even dimensions and associated with tensor fields. A tensor monopole in 4D space $\boldsymbol q=(q_x,q_y,q_z,q_z)$ takes a (3-form) curvature tensor $\mathcal{H}_{\mu\nu\lambda}$ \cite{Palumbo2018,Palumbo2019}, as the generalization of the (2-form) Berry curvature $\mathcal{F}_{\mu\nu}$ of the Dirac monopole. A minimal model realizing such a tensor monopole is the three-band Weyl-like Hamiltonian in 4D space \cite{Palumbo2018}:
\begin{eqnarray}
\begin{aligned}
H_{4D}=\boldsymbol q\cdot\boldsymbol{\lambda}=\begin{bmatrix}
0&q_{x}-i q_{y}&0 \\
q_{x}+i q_{y}&0&q_{z}+i q_{w} \\
0&q_{z}-i q_{w}&0
\end{bmatrix},
\end{aligned}
\label{Ham_4D}
\end{eqnarray}
where $\mathbf{\lambda}=(\lambda_1,\lambda_2,\lambda_6,\lambda_7^{\ast})$ are $3\times3$ Gell-Mann matrices. The energy spectrum is given by $E_{0,\pm}=0,\pm|\boldsymbol{q}|$, with a triple-degenerate Weyl-like point at $\boldsymbol{q}=(0,0,0,0)$ in 4D parameter space. Such a Weyl-like node gives a tensor monopole, surrounded by a 3D hypersphere $S^3$. In terms of hyperspherical coordinates $\{r,\theta_1,\theta_2,\phi \}~$ ($\theta_{1,2}\in[0,\pi]$ and $\phi\in [0,2\pi]$), one has $q_x=r\cos\theta_1$, $q_y=r\sin\theta_1\cos\theta_2$, $q_z=r\sin\theta_1\sin\theta_2\cos\phi$, and $q_w=r\sin\theta_1\sin\theta_2\sin\phi$. The generalized curvature tensor as the field strength in $S^3$ is related to the quantum metric \cite{Palumbo2018}:
\begin{equation}
 \mathcal{H}_{\theta_1\theta_2\phi}=\epsilon_{\theta_1\theta_2\phi} (4\sqrt{\det g_{\mu \nu}}), ~~\mu,\nu=\{\theta_1,\theta_2,\phi\}.
 \label{tensorfield}
 \end{equation}
Here $H_{4D}$ has $\phi$-rotation symmetry and thus $\mathcal{H}_{\theta_1\theta_2\phi}$ is independent of $\phi$. For the ground state $|\psi_{-}\rangle$ of the system, 
all matrix elements of the metric tensor $g$ can be explicitly obtained (see Eqs. (S7) in SM \cite{Supp}). The tensor monopole generalizes the Dirac monopole to 4D, and takes a topological charge associated with the generalized curvature tensor $\mathcal{H}_{\theta_1\theta_2\phi}$:
\begin{eqnarray}
Q_{DD}=\frac{1}{2\pi^2}\int_{0}^{\pi}d\theta_1\int_{0}^{\pi}d\theta_2\int_{0}^{2\pi}d\phi \mathcal{H}_{\theta_1\theta_2\phi}=1, \label{Qt}
\end{eqnarray}
which is the DD invariant \cite{Mathai,Murray}. Thus, to obtain the topological charge $Q_{DD}$ of a tensor monopole, one can measure $\mathcal{H}_{\theta_1\theta_2\phi}$ by revealing the quantum metric $g_{\mu\nu}$.

In parameter space, the quantum distance $ds^2$ is related to the transition probability $P^+$ of the quantum state being excited to other eigenstates after a sudden quench: $P^+=ds^2$ \cite{Kolodrubetz,Lim2015,Tan2019b}. One can thus measure the quantum metric via  transition probability by the sudden quench method. For a quantum state initially prepared at $\boldsymbol q$, to extract the diagonal components $g_{\mu\mu}$ at this point, one can suddenly quench the system parameter to $\boldsymbol q+\delta q \boldsymbol e_{\mu}$ along the $\boldsymbol e_{\mu}$ direction, and then measure the transition probability $P^+_{\mu\mu}=g_{\mu\mu}\delta q^2+\mathcal{O}(\delta q^3)$. To extract the off-diagonal components $g_{\mu\nu}$ ($\mu\neq\nu$), we apply a sudden quench to $\boldsymbol q+\delta q \boldsymbol e_{\mu}+\delta q\boldsymbol e_{\nu}$ along the $\boldsymbol e_{\mu}+\boldsymbol e_{\nu}$ direction and then measure the probability $P^+_{\mu\nu}$, which has the relation $P^+_{\mu\nu}-P^+_{\mu\mu}-P^+_{\nu\nu}=2g_{\mu\nu}\delta q^2+\mathcal{O}(\delta q^3)$. This sudden quench scheme 
will be used to measure the quantum metric $g_{\mu\nu}$ in Eq. (\ref{tensorfield}).

\begin{figure}[tbph]
	\includegraphics[width=7.5cm]{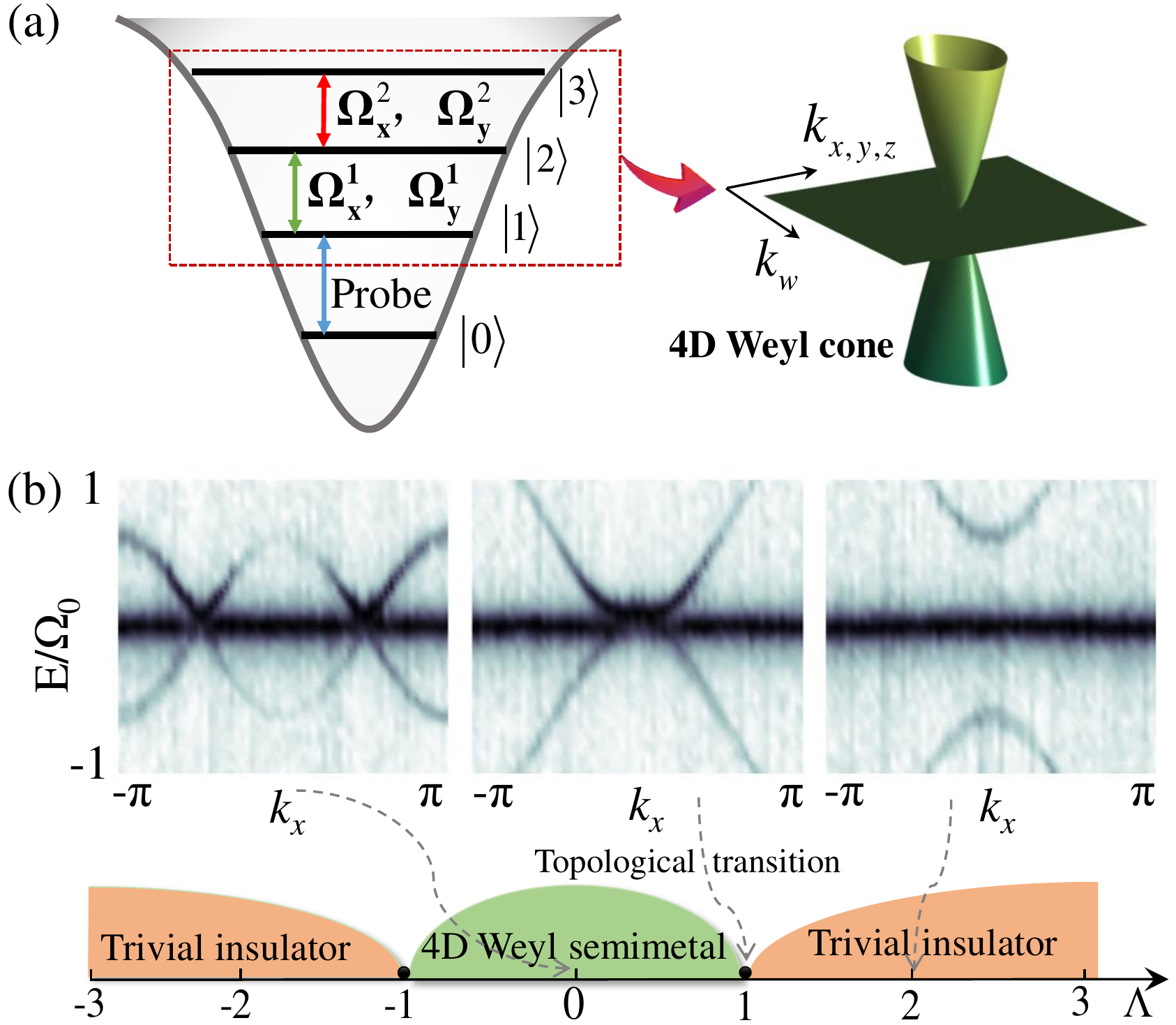}
	\caption{(Color online) Measurement of the energy structure of a 4D Weyl-like semimetal. (a) Diagram of energy levels in superconducting circuit. $|1\rangle$, $|2\rangle$ and $|3\rangle$ are used to construct the Hamiltonian with irradiated microwaves, while $|0\rangle$ is for detecting the spectrum. (b) Measured energy structure with different offsets $\Lambda=0,1,2$ in the phase diagram. }
	\label{fig2}
\end{figure}

\emph{Experimental system.}--We realize a highly tunable spin-1 Hamiltonian with superconducting quantum circuits and observe the energy spectrum and topological charge of the tensor monopole in parameter space. The circuits consist of a superconducting transmon qubit embedded in a 3D aluminum cavity~\cite{Tan2018,Tan2019a,paik_3d,devoret_3d,JinXY,DiCarlo}.
The resonance frequency of the cavity TE101 mode is 9.0526 GHz. The whole sample package is cooled in a dilution refrigerator to a base temperature of 20 mK. The experimental setup for the qubit control and measurement is well established \cite{Tan2018,Tan2019a,paik_3d,devoret_3d,JinXY,DiCarlo}.
The coupled transmon qubit and cavity exhibit anharmonic multiple energy levels. In our experiments, the lowest four energy levels $|0\rangle $, $|1\rangle $, $|2\rangle $ and $|3\rangle$ are used and form a qudit system shown in Fig. \ref{fig2}(a). Among them, three levels are chosen to construct the Hamiltonian in Eq. (\ref{Ham_exp}), which are $\{1,2,3\}$ and $\{0,1,2\}$ for measurements of energy structures and quantum metric, respectively \cite{Supp}. Microwave fields are applied to couple the four energy levels. The transition frequencies between them are $\omega_{10}/2\pi =$ 7.1194 GHz, $\omega_{12}/2\pi =$ 6.7747 GHz and $\omega_{23}/2\pi =$ 6.3926 GHz respectively, which are independently determined by saturation spectroscopy \cite{Supp}.
We apply microwave driving along $x$, $y$, and $z$ directions and realize the following  effective Hamiltonian in the rotating frame ($\hbar =1$) \cite{Supp}

\begin{eqnarray}
\begin{aligned}
H_{\text{exp}}
=\frac{1}{2}\begin{bmatrix}
0&\Omega^{1}_{x}-i \Omega^{1}_{y}&0 \\
\Omega^{1}_{x}+i \Omega^{1}_{y}&0&\Omega^{2}_{x}+i \Omega^{2}_{y} \\
0&\Omega^{2}_{x}-i \Omega^{2}_{y}&0
\end{bmatrix},
\end{aligned}
\label{Ham_exp}
\end{eqnarray}
where $\Omega^{1(2)}_{x}$ $(\Omega^{1(2)}_{y})$ is the Rabi frequency along the $x$ ($y$) axis of the Bloch sphere spanned by the corresponding basis. For the case shown in Fig. \ref{fig2}(a), the system parameters $\Omega^{1}_{x,y}$ [$(\Omega^{2}_{x,y})$] are fully controlled by the amplitude and phase of the microwave applied to couple $|1\rangle$ and $|2\rangle$ ($|2\rangle$ and $|3\rangle$). By varying these parameters, we can create arbitrary three-level Hamiltonians given by Eq. (\ref{Ham_exp}). In our experiments, we work with collections of Hamiltonians represented in the 4D parameter space by accurately designing microwave fields after calibration of the parameters using Rabi oscillations and Ramsey fringes \cite{Supp}.

\emph{Measuring energy structures of 4D Weyl model.}--We obtain the energy
structure by measuring the spectrum of the qudit system. After mapping the momentum space of a 4D Weyl-semimetal Hamiltonian \cite{Palumbo2019,Supp} to the parameter space of the system Hamiltonian in Eq. (\ref{Ham_exp}), we can visualize the simulated energy structures. We design the Rabi frequencies $\{\Omega^{1}_{x},\Omega^{1}_{y},\Omega^{2}_{x},\Omega^{2}_{y}\}=\{\Omega_0 (3+\Lambda -\cos k_x -\cos k_y -\cos k_z -\cos k_w, \Omega_0 \sin k_y, \Omega_0 \sin k_z, \Omega_0 \sin k_w \}$, where $\Omega_0=5$ MHz is the energy unit and the parameter $\Lambda$ is added to account for an offset in $\Omega^{1}_{x}$. As shown in Fig. \ref{fig2}(a), the energy levels $\{|1\rangle,|2\rangle,|3\rangle\}$ are used to construct $H_{\text{exp}}$ and $|0\rangle$ is treated as a reference level for spectrum probing. The dressed states under the coupled microwaves are eigenstates of the Hamiltonian (\ref{Ham_exp}) labelled $|\psi_0\rangle$ and $|\psi_{\pm}\rangle$. Notably, the fictitious momenta $k_{x,y,z,w}$ (the indexes $x,y,z$ are not related to real spatial coordinates of the experimental system) denote the 4D parameter space controlled by varying $\Omega^{1,2}_{x,y}$ in our system and $\Lambda$ plays the role of a fictitious Zeeman field for tuning topological phase transition \cite{Palumbo2019,Supp}. Similar mapping procedures were used to simulate other condensed-matter models in engineered quantum systems \cite{Schroer2014,Roushan2014,Tan2018,Tan2019a,Sugawa}.

\begin{figure}[tbph]
	\includegraphics[width=8cm]{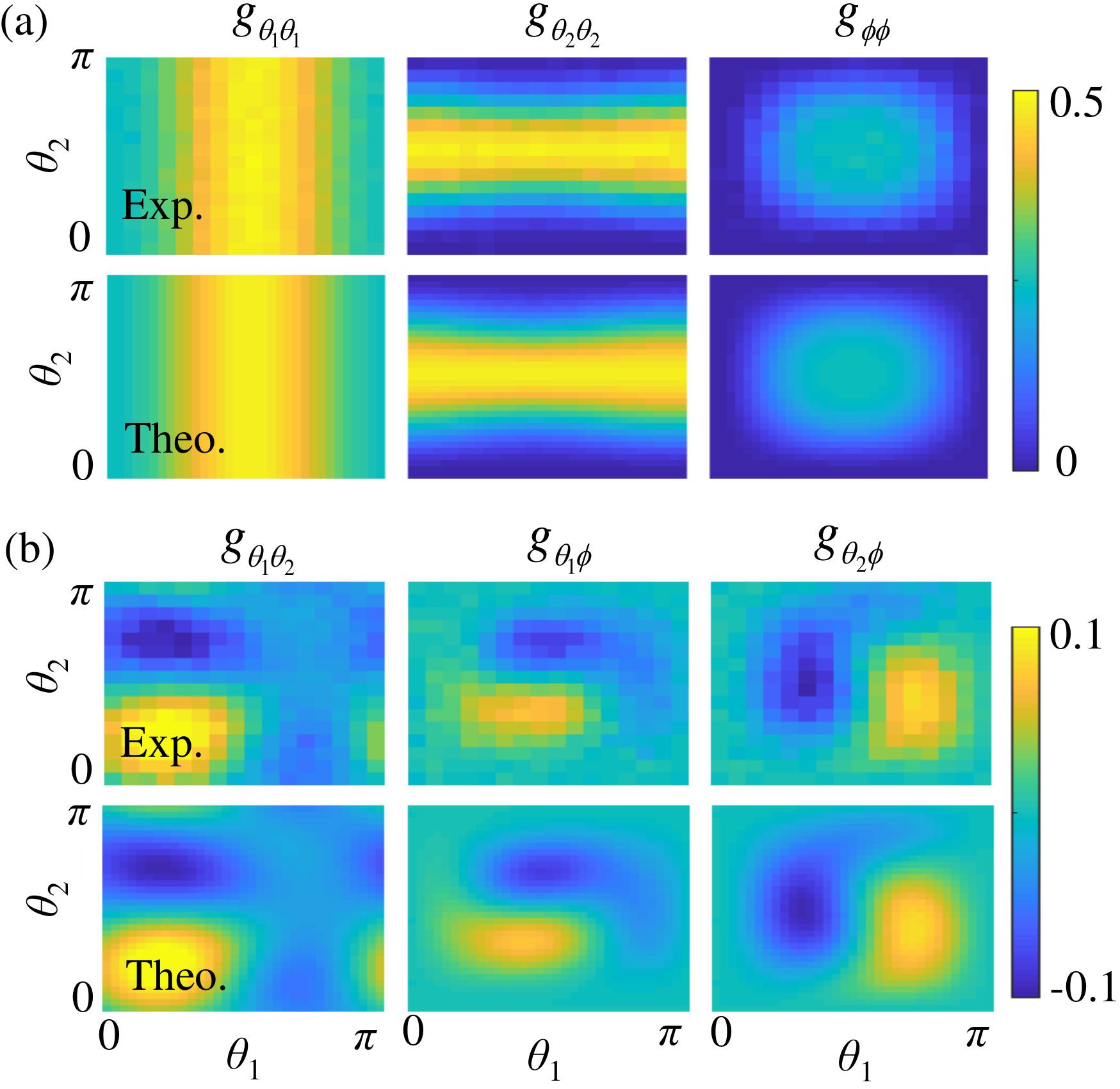}
	\caption{(Color online) Experimental and theoretical results of the quantum metric $g_{\mu \nu}$ as a function of  $\theta_1$ and $\theta_2$ for (a) diagonal components; and (b) off-diagonal components.}
	\label{fig3}
\end{figure}

In our routine, we execute the spectrum-like measurement and the resonant peaks of microwave absorption are detected \cite{Supp}. The frequency of the resonant peak is a function of $k_{x,y,z,w}$, and we are able to extract the energy structure of the 4D Weyl-like cone, as illustrated in the right panel of Fig. \ref{fig2}(a). To demonstrate the topological properties, we set $k_{y,z,w}=0$ to emphasize the $E$-$k_{x}$ plane, where the phase transition can be clearly observed. The system has two different phases determined by the parameter $\Lambda$, as shown in Fig. \ref{fig2}(b): the 4D Weyl-like semimetal with a pair of 4D Weyl points when $|\Lambda|<1$ and the trivial gapped insulator when $|\Lambda|>1$ \cite{Palumbo2018,Palumbo2019}. At the critical points $|\Lambda|=1$, two degenerate points merge and then disappear. The extracted energy structures for $\Lambda=0,1,2$ are illustrated in Fig. \ref{fig2}(b), which capture the features of the theoretical prediction with two degenerate points at $K_{\pm}=(\pm\pi/2,0,0,0)$ when $\Lambda=0$. Near $K_{\pm}$, one has
the low-energy effective Hamiltonian $H_{4D}^{\pm}=\pm q_x\lambda_1+q_y\lambda_2+q_z\lambda_6+q_w\lambda_7^{\ast}$ describing a pair of tensor monopoles with $Q_{DD}=\pm1$ \cite{Supp}, where the sign in front of $q_x$ determines the topological charges. Below we focus on the tensor monopole described by $H_{4D}^{+}=H_{4D}$ in Eq. (\ref{Ham_4D}) and $Q_{DD}=1$.

\begin{figure}[tbph]
	\includegraphics[width=7.5cm]{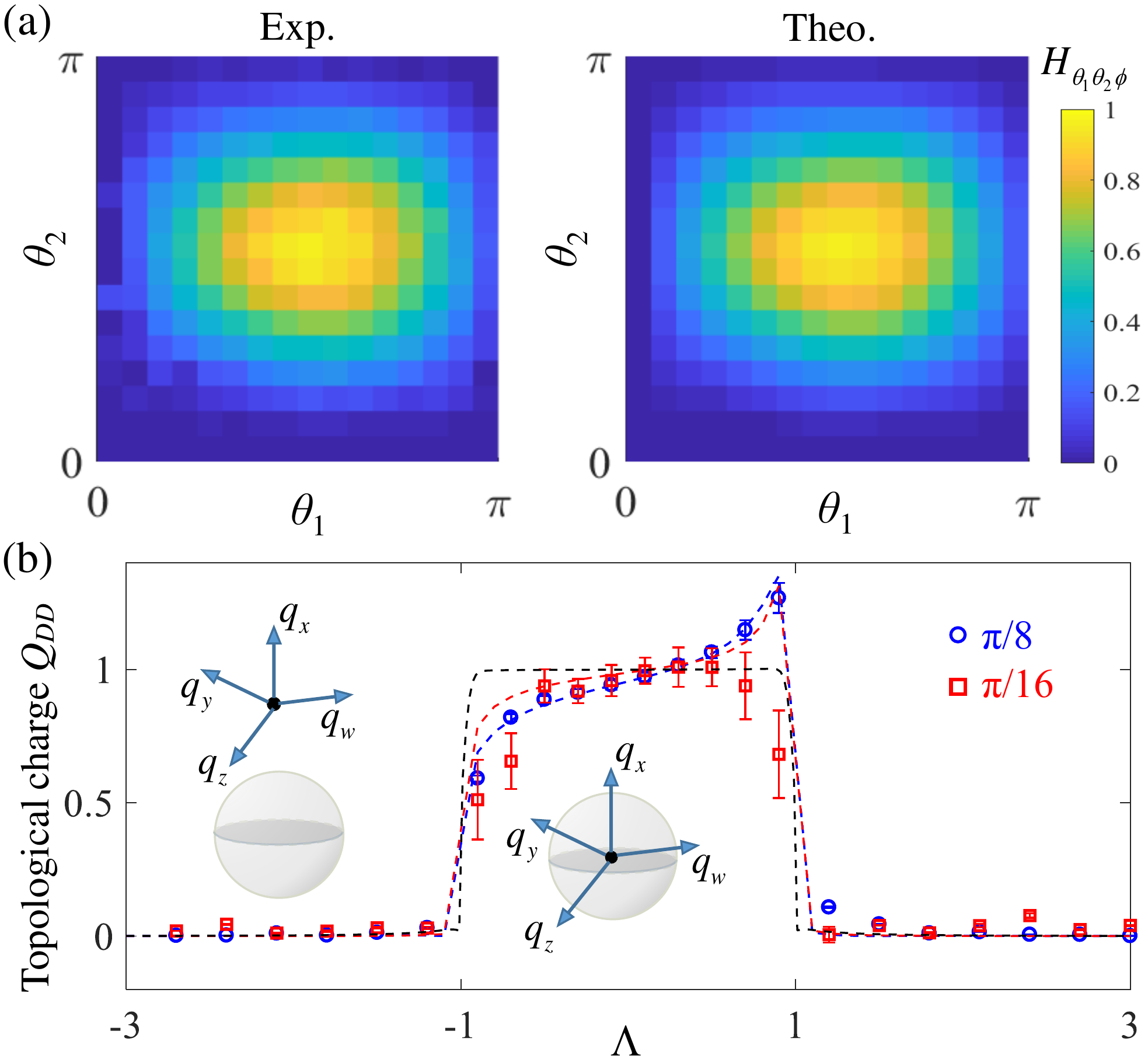}
	\caption{(Color online) Topological phase transition Characterized by tensor monopole charge (a) Experimental and theoretical results of generalized curvature tensor $\mathcal{H}_{\theta_1\theta_2\phi}$ as a function of $\theta_1$ and $\theta_2$. (b) Topological charge $Q_{DD}$ as a function of offset $\Lambda$. $Q_{DD}\approx1$ declines rapidly to $Q_{DD}\approx0$ when $|\Lambda|>1$ with the tensor monopole outside the $S^3$ sphere, indicating topological transition. Data obtained with $\delta q=\pi/8$, and $\pi/16$ are shown in blue and red, respectively, with symbols and dashed lines respresenting experimental data and numerical simulations, respectively. Black dashed line is the simulation result for $\delta q=\pi/1024$. Some deviations between the experimental and simulation results for $\delta q=\pi/16$ is due to the reduction of the measurement accuracy of the excitation probability in our quench scheme for smaller $\delta q$.}
	\label{fig4}
\end{figure}

\emph{Measuring quantum metric by sudden quench.}--We now measure the quantum metric $g_{\mu\nu}$ ($\mu,\nu$=$\{\theta_1,\theta_2,\phi\}$) of the simulated tensor monopole using the sudden quench scheme. We here work with the three lowest-energy levels $\{|0\rangle$,$|1\rangle$,$|2\rangle\}$ without a reference level since the
spectrum probing is unnecessary \cite{Supp}. We construct the Hamiltonian in hyper-sphere coordinates with parameters in Eq. (\ref{Ham_exp}) as $\{\Omega^{1}_{x}= \Omega_0\cos\theta_1,\Omega^{1}_{y}= \Omega_0\sin\theta_1\cos\theta_2,\Omega^{2}_{x}= \Omega_0\sin\theta_1\sin\theta_2\cos\phi,\Omega^{2}_{y}= \Omega_0\sin\theta_1\sin\theta_2\sin\phi\}$ \cite{Supp}. The system is initially prepared in the ground state $|\psi_{-}\rangle$ in the parameter space $\boldsymbol q=\{\theta_1,\theta_2,\phi\}$ with $\phi=0$. The Hamiltonian is then rapidly swept to $H(\boldsymbol q+\delta\boldsymbol q)$, followed by state tomography to obtain the transition probability. We set the quench parameter to $\boldsymbol q(t)=\boldsymbol q+t/T\delta q\boldsymbol e$ along the $\boldsymbol e$ direction, where the quench time $T=9$ ns and $\delta q=\pi/8$ or $\pi/16$ \cite{Supp}. 
For the diagonal term $g_{\mu\mu}$, only one parameter ramps linearly in each quench with $\boldsymbol e=\{\boldsymbol e_{\theta_1}, \boldsymbol e_{\theta_2},\boldsymbol e_{\phi}\}$, respectively. For the off-diagonal term $g_{\mu\nu}$ ($\mu\neq\nu$), the parameters  $\mu$ and $\nu$ ramp simultaneously, with $\boldsymbol e=\{\boldsymbol e_{\theta_1}+\boldsymbol e_{\theta_2},
\boldsymbol e_{\theta_1}+\boldsymbol e_{\phi},\boldsymbol e_{\theta_2}+\boldsymbol e_{\phi}\}$. These ramp procedures are illustrated in Fig. \ref{fig1}(b). From the final state's tomography, we extract the metric at $\boldsymbol q$ from the measured transition probability: $g_{\mu\mu}\approx P_{\mu\mu}/\delta q^2$ and $g_{\mu \nu}\approx(P_{\mu \nu}-P_{\mu\mu}-P_{\nu \nu})/2\delta q^2$. The measured $g_{\mu\nu}$ as a function of $\theta_1$ and $\theta_2$  are shown in Fig. \ref{fig3}, which agree well with theoretical results.

\emph{Observing topological phase transitions.}-- To further study the tensor monopole, we observe topological phase transition characterized by the tensor monopole charge in our superconducting circuits. By designing microwave fields on the qudit, we modify Eq. (\ref{Ham_exp}) by adding a tunable offset $\Lambda$ into the $\Omega^{1}_x$ term, such that $\Omega^{1}_x=\Omega_0(\cos\theta_1+\Lambda)$, while other terms remain unchanged (without breaking the $\phi$-rotation symmetry). By measuring the metric tensor with the sudden-quench approach, we can obtain the generalized curvature $\mathcal{H}_{\theta_1\theta_2\phi}$ and then integrate it to derive the topological charge $Q_{DD}$. For offset $\Lambda=0$, the extracted $\mathcal{H}_{\theta_1\theta_2\phi}$ as a function of  parameters $\theta_1$ and $\theta_2$ is shown in Fig. \ref{fig4}(a). Experimental data (left) agree with theoretical results (right). We finally calculate the $Q_{DD}$ using Eq. (\ref{Qt}) and obtain $Q_{DD}=0.92\pm0.15$ for $\Lambda$=0.

To study the topological phase transition, we execute the protocol with varying $\Lambda$. The extracted DD invariant as a function of $\Lambda$ is shown in Fig. \ref{fig4}(b).  When $|\Lambda|=0$, the manifold of the parameter space $S^3$ surrounds the tensor monopole in the center. With the increase of $|\Lambda|$, the tensor monopole moves along the $q_x$ axis. $Q_{DD}\approx1$ when $|\Lambda|<1$ for the $S^3$ sphere surrounding the tensor monopole. $Q_{DD} \approx 0$  when $|\Lambda|>1$ since the monopole moves outside the hyper-sphere manifold, indicating that the system is in the trivial insulator phase. $Q_{DD}$ declines rapidly to around $0$ in the vicinity of $\Lambda=\pm1$, which indicates a topological phase transition. The accuracy of the topological charge extracting from the sudden quench routine depends on the ramp step. In Fig. \ref{fig4}(b), the numerical results with $\delta q$=$\pi/1024$ are plotted, which are very close to the expected integer values. However, such a small step is not feasible to implement in practice due to limitation of readout fidelity. With a larger $\delta q$, measurement obtained from the sudden quench routine will deviate from ideal values. For comparison, we perform the routine with $\delta q=\pi/8$ and $\pi/16$, as demonstrated in Fig. \ref{fig4}(b). When $\delta q$ decreases, the deviation from the ideal quantized values becomes smaller.

\emph{Conclusion.}--In summary, we have created tensor monopoles in 4D parameter space and explored their unique properties  using superconducting circuits. Our experimental observation contributes to exploring tensor gauge fields in quantum mechanics and
creates a unique approach in the search for exotic topological matter in condensed matter physics and artificial
systems, such as topological semimetals and unconventional quasiparticles beyond Dirac and Weyl fermions in high dimensions. By coupling individual superconducting qudits, one can further explore the geometric and topological properties of quantum many-body systems.


\textit{Note added:} After our submission, we noticed another work on experimental observation of the tensor monopole using a single
nitrogen-vacancy center in diamond \cite{MChen}.

\acknowledgments
This work was supported by the National Key Research and Development Program of China (Grant No. 2016YFA0301800), the NNSF of China (Grants No. 11474153, No. 91636218, No.11890704, No. 61521001, and No. 12074180 ), the NSAF (Grant No. U1830111), the Key-Area Research and Development Program of Guangdong Province (Grants No. 2018B030326001 and No. 2019B030330001), and the Key Project of Science and Technology of Guangzhou (Grants No. 201804020055 and No. 2019050001).

X. T. and D.-W. Z contributed equally to this work.

\clearpage

\onecolumngrid
\appendix

\section{Supplemental Materials}

\subsection{Quantum metric and topological charge of a tensor monopole}
	
For a general Hamiltonian $H(\boldsymbol q)$ in the parameter space $\boldsymbol q=(q_1, q_2, \cdots, q_N)\in \mathcal{M}$, one has the eigen-energies $E_n(\boldsymbol q)$ and eigen-states $|u_n(\boldsymbol q)\rangle$ at each point of the manifold $\mathcal{M}$. In the absence of energy degeneracies, the quantum geometric tensor associated with $|u_{\boldsymbol q}\rangle$ is defined as \cite{Kolodrubetz,Provost}
	\begin{equation}\label{QGT}
	\chi_{\mu\nu}=\langle \partial_{q_{\mu}} u_{\boldsymbol q} | \partial_{q_{\nu}}u_{\boldsymbol q} \rangle-\langle \partial_{q_{\mu}} u_{\boldsymbol q} | u_{\boldsymbol q}\rangle \langle u_{\boldsymbol q}| \partial_{q_{\nu}}u_{\boldsymbol q} \rangle=\langle \partial_{q_{\mu}} u_{\boldsymbol q} |(1 - | u_{\boldsymbol q}\rangle \langle u_{\boldsymbol q}|)| \partial_{q_{\nu}}u_{\boldsymbol q} \rangle.
	\end{equation}
A generalized quantum geometric tensor can be defined for degenerate states \cite{Ma2010,Rezakhani}. The real part of this geometric tensor is symmetric and defines the quantum metric $g_{\mu\nu} = \mathrm{Re}[\chi_{\mu\nu}]=g_{\nu\mu}$, which is the so-called Fubini-Study metric on the projective Hilbert space $\mathcal{P}H(\boldsymbol q)=H(\boldsymbol q)/U(1)$ (here we identify quantum states differ only by a local phase factor), required by the principle of gauge invariance \cite{Provost}. The imaginary part is related to the well-known anti-symmetric Berry curvature $\mathcal{F}_{\mu\nu} =-2\mathrm{Im}[\chi_{\mu\nu}]=-\mathcal{F}_{\nu\mu}$. The quantum metric $g_{\mu\nu}$ measures the quantum distance between nearby states $|u_{\boldsymbol q}\rangle$ and $|u_{\boldsymbol q+\delta\boldsymbol q}\rangle$ as
	\begin{equation}\label{ds}
	ds^2=P^{+}=1-|\langle u_{\boldsymbol q}| u_{\boldsymbol q+\delta\boldsymbol q}\rangle|^2=\sum_{\mu\nu} \chi_{\mu\nu} \delta q_{\mu} \delta q_{\nu}+\mathcal{O}(|\delta\boldsymbol q|^3)=\sum_{\mu\nu} g_{\mu\nu} \delta q_{\mu} \delta q_{\nu}+\mathcal{O}(|\delta\boldsymbol q|^3),
	\end{equation}
where $ds^2$ is determined by the wave-function overlap (a maximal overlap of 1 corresponds to the zero distance $ds^2=0$, while the orthogonal states correspond to the maximal distance $ds^2=1$), and $P^{+}$ is the probability to excite the system to other eigen-states after a quench with the parameters suddenly changing from $\boldsymbol q$ to $\boldsymbol q+\delta\boldsymbol q$ \cite{Tan2019b}.

For the 4D Weyl semimetals in the momentum space $\boldsymbol k=(k_x,k_y,k_z,k_w)$ ($k_{x,y,z,w}\in[-\pi,\pi]$), one has the three-band Bloch Hamiltonian \cite{Palumbo2018,Palumbo2019}
\begin{eqnarray}
\begin{aligned}
H_k=\begin{bmatrix}
0&d_x-i \sin k_{y}&0 \\
d_x+i \sin k_{y}&0&\sin k_{z}+i \sin k_{w} \\
0&\sin k_{z}-i \sin k_{w}&0
\end{bmatrix}, ~~ d_x=3+\Lambda -\cos k_x -\cos k_y -\cos k_z -\cos k_w
\end{aligned},
\label{Ham1}
\end{eqnarray}
where $\Lambda$ is a parameter for tuning the topological phase transition. When $|\Lambda|<1$, there is a pair of 4D Weyl points located at $\boldsymbol K_{\pm}=(\pm \arccos\Lambda,0,0,0)$ and the model describes a 4D topological Weyl-like semimetal. When $|\Lambda|>1$, the model describes a trivial insulator with the topological phase transition at $|\Lambda|=1$. Near the points $\boldsymbol K_{\pm}$, one has the low-energy Weyl-like Hamiltonian \cite{Palumbo2018}:

\begin{eqnarray}
\begin{aligned}
H_{4D}^{\pm}=\pm q_x\lambda_1+q_y\lambda_2+q_z\lambda_6+q_w\lambda_7^{\ast}=\begin{bmatrix}
0&\pm q_{x}-i q_{y}&0 \\
\pm q_{x}+i q_{y}&0&q_{z}+i q_{w} \\
0&q_{z}-i q_{w}&0
\end{bmatrix},
\end{aligned}
\label{Ham_4D}
\end{eqnarray}
with $\boldsymbol q=\boldsymbol k-\boldsymbol K_{\pm}$. Note that $H_{4D}^{\pm}$ describe the monopole and anti-monopole with opposite topological charges [i.e., $Q_{DD}=\pm1$ given by Eq. (\ref{Qt})], respectively. In our experiments, we focus on the monopole with the Hamiltonian $H_{4D}^{+}\equiv H_{4D}$ without loss of generality. Similar as those in Refs. \cite{Schroer2014,Roushan2014,Tan2018,Tan2019a}, we can map the momentum space of the condensed-matter models $H_k$ and $H_{4D}$ to the parameter space of three-level Hamiltonians.  Note that hereafter and in the main text, we still take $\boldsymbol k$ and $\boldsymbol q$ to denote the corresponding 4D parameter space by for simplicity.

For the Hamiltonian $H_{4D}$, the energy  spectra are given by $E_{0,\pm}=0,\pm|\boldsymbol{q}|$ for three eigenstates $|\psi_0\rangle$ and $|\psi_{\pm}\rangle$, respectively. At $\boldsymbol{q}=(0,0,0,0)$ in 4D parameter space, a triple-degenerate Weyl-like point acts as a tensor monopole, which is surrounded by a 3D hypersphere $S^3$. In terms of hyperspherical coordinates $\{r,\theta_1,\theta_2,\phi \}~$ ($\theta_{1,2}\in[0,\pi]$ and $\phi\in [0,2\pi]$), one has
\begin{eqnarray}
\begin{aligned}
q_x&=r\cos\theta_1,\\
q_y&=r\sin\theta_1\cos\theta_2, \\
q_z&=r\sin\theta_1\sin\theta_2\cos\phi,\\
q_w&=r\sin\theta_1\sin\theta_2\sin\phi,
\end{aligned}
\end{eqnarray}
in Eq. (\ref{Ham_4D}). We consider the ground state $|\psi_{-}\rangle=(\cos\theta_1-i \cos\theta_2\sin\theta_1,-1,\sin\theta_1\sin\theta_2 e^{-i\phi})^\text{T}$ with $T$ denoting the transposition of matrix, the $3\times3$ quantum metric tensor $g$ in the $S^3$ is given by
\begin{eqnarray}
\begin{aligned}
g=\begin{bmatrix}
g_{\theta_1\theta_1}&g_{\theta_1\theta_2}&g_{\theta_1\phi} \\
g_{\theta_2\theta_1}&g_{\theta_2\theta_2}&g_{\theta_2\phi} \\
g_{\phi\theta_1}&g_{\phi\theta_2}&g_{\phi\phi}
\end{bmatrix},
\end{aligned}
\end{eqnarray}
where the three diagonal components and six off-diagonal components are derived as
\begin{eqnarray}
\begin{aligned}
g_{\theta_1\theta_1}&=\frac{1}{8}(3-\cos 2\theta_1), \\ g_{\theta_2\theta_2}&=\frac{1}{4}\sin^2\theta_1[2\cos^2\theta_2-(\cos^2\theta_1-2)\sin^2\theta_2],\\ g_{\phi\phi}&=-\frac{1}{4}\sin^2\theta_1\sin^2\theta_2(\sin^2\theta_1\sin^2\theta_2-2),\\ g_{\theta_1\theta_2}&=g_{\theta_2\theta_1}=\frac{1}{4}\cos\theta_1\cos\theta_2\sin\theta_1\sin\theta_2, \\ g_{\theta_1\phi}&=g_{\phi\theta_1}=-\frac{1}{4}(\cos\theta_2\sin^2\theta_1\sin^2\theta_2), \\
g_{\theta_2\phi}&=g_{\phi\theta_2}= \frac{1}{4}\cos\theta_1\sin^2\theta_1\sin^3\theta_2.
\end{aligned}
\end{eqnarray}
It has been shown that the quantum metric is related to the generalized curvature tensor $\mathcal{H}_{\theta_1\theta_2\phi}$ as the field strength of the tensor monopole in $S^3$ \cite{Palumbo2018,Palumbo2019}:
\begin{equation}
 \mathcal{H}_{\theta_1\theta_2\phi}=\epsilon_{\theta_1\theta_2\phi} (4\sqrt{\det g_{\mu \nu}}), ~~\mu,\nu=\{\theta_1,\theta_2,\phi\}.
 \end{equation}
Here $\mathcal{H}_{\theta_1\theta_2\phi}$ is independent on $\phi$ as $H_{4D}$ has $\phi$-rotation symmetry. The tensor monopole has a topological charge (the DD invariant) associated with the curvature $\mathcal{H}_{\theta_1\theta_2\phi}$:
\begin{eqnarray}
Q_{DD}=\frac{1}{2\pi^2}\int_{0}^{\pi}d\theta_1\int_{0}^{\pi}d\theta_2\int_{0}^{2\pi}d\phi \mathcal{H}_{\theta_1\theta_2\phi}=\frac{1}{2\pi^2}\int_{0}^{\pi}d\theta_1\int_{0}^{\pi}d\theta_2\int_{0}^{2\pi}d\phi \sin^2\theta_1\sin\theta_2=1. \label{Qt}
\end{eqnarray}
For $H_{4D}^{-}$ in Eq. (\ref{Ham_4D}), the corresponding topological charge can be obtained as $Q_{DD}=-1$.

\begin{figure}[tbph]
\includegraphics[width=10cm]{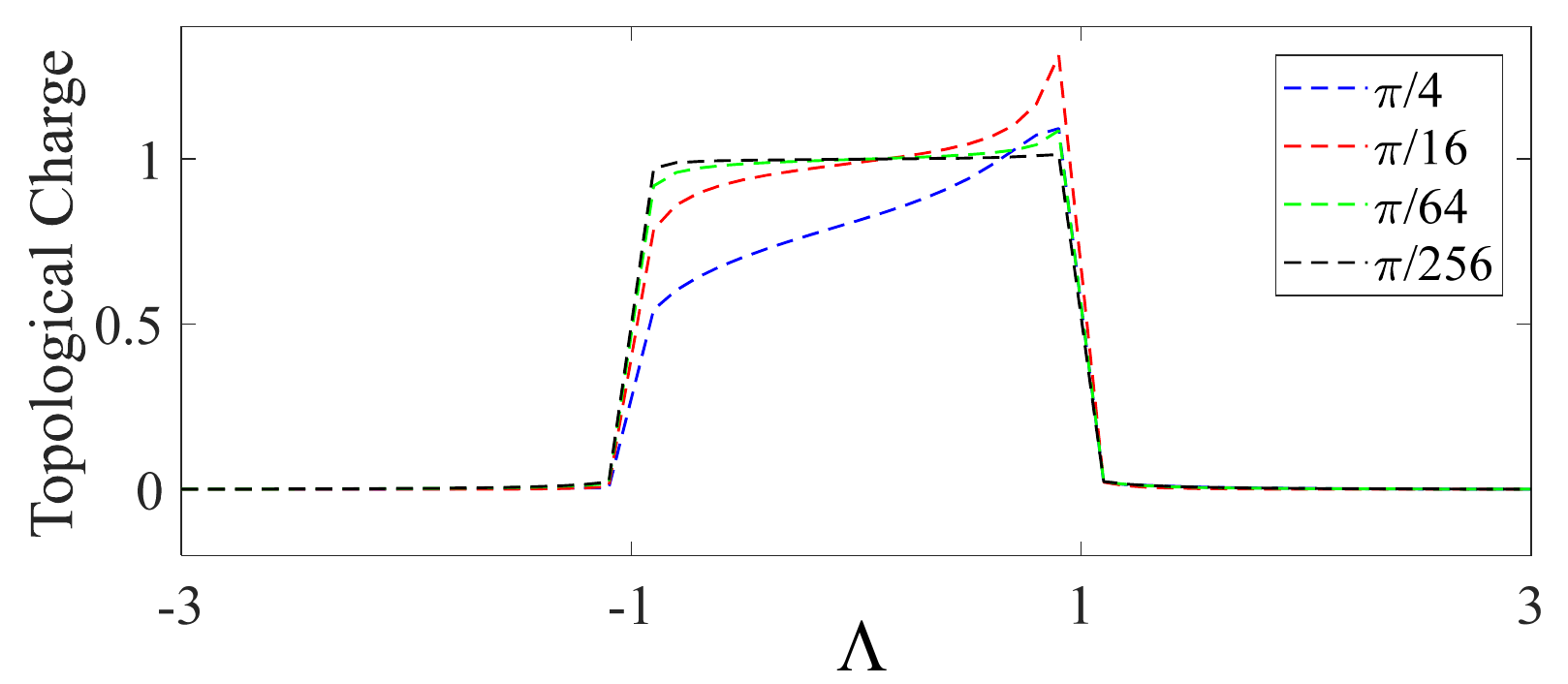}
\caption{Simulation results for the topological charge $Q_{DD}$ as a function of offset $\Lambda$ for different $\delta q$. }
\label{s1}	
\end{figure}

Thus, by revealing the quantum metric $g_{\mu\nu}$ through the sudden quench scheme, we can measure $\mathcal{H}_{\theta_1\theta_2\phi}$ and then obtain the topological charge $Q_{DD}$ of the tensor monopole. See the main text for the ramp procedures. The obtained numerical results of $Q_{DD}$ as a function of an additional offset $\Lambda$ for varying ramp step $\delta q$ are shown in Fig. \ref{s1}. When $\delta q$ decreases, the deviation from the ideal quantized values ($Q_{DD}=0$ when $|\Lambda|>1$ and $Q_{DD}=1$ when $|\Lambda|<1$) becomes smaller. For $\delta q$=$\pi/256$, the obtained $Q_{DD}$ is very close to the integer values.

	\subsection{Experimental setup and qubit calibration}
	
\begin{figure}[tbph]
	\includegraphics[width=14cm]{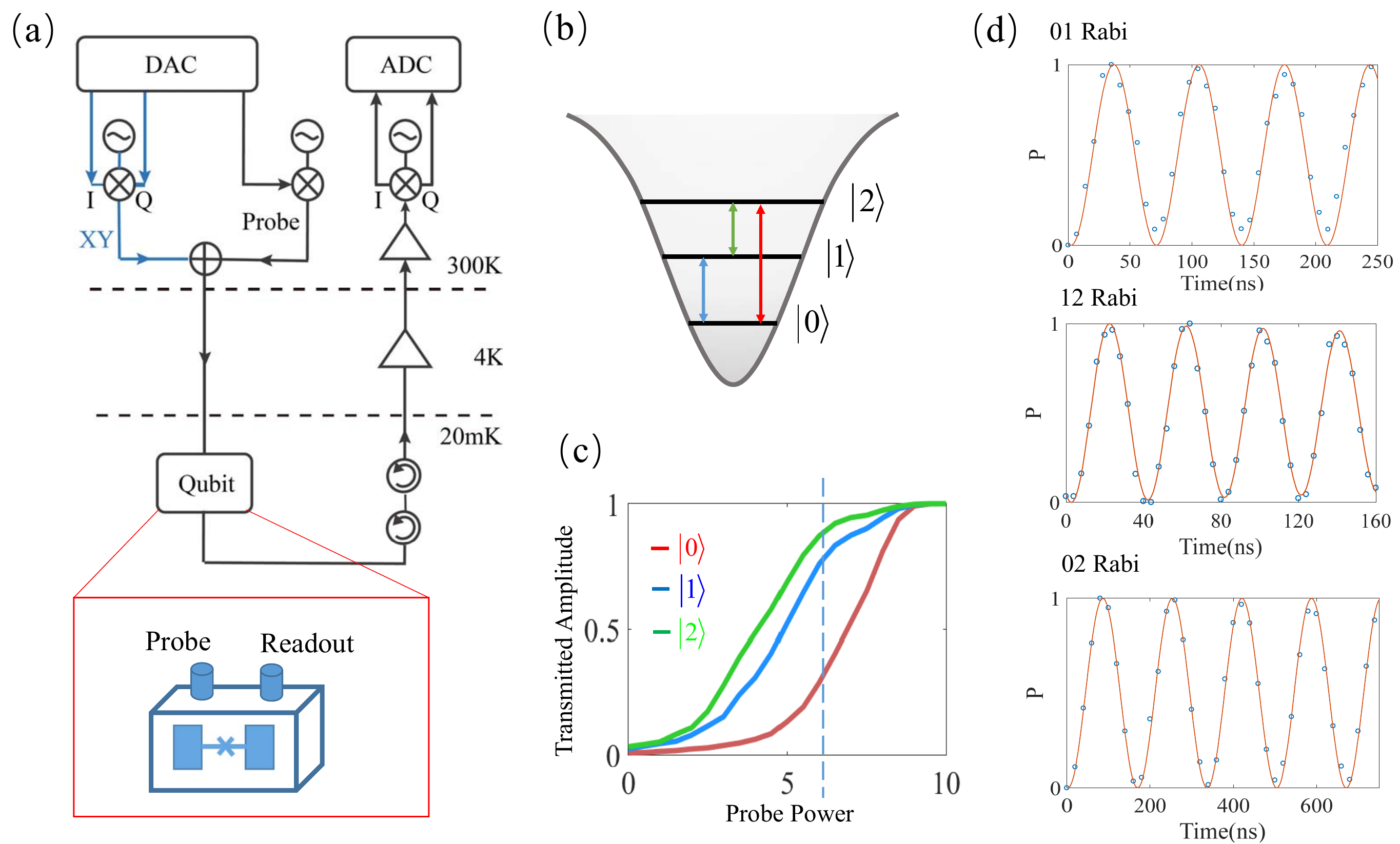}
	\caption{(a) Schematic of the experimental setup. 3D transmon is cooled down in dilution refrigerator with base temperature 20 mk. Circulator and amplifiers are applied to increase the signal-to-noise ratio. The qubit state is read out though cavity with the Heterodyne technique. The shunting pads of sample form a dipole antenna which couples the qubit to the cavity with strength $g/2\pi$ = 280 MHz. (b) Visualization of energy levels used in measurement of metric tensor. Here, the lowest three levels are labelled as $|0\rangle$, $|1\rangle$ and $|2\rangle$. Transition between these states are pre-calibrated, while the corresponding Rabi oscillation are demonstrated in (d). (c) Transmitted amplitude of cavity signal as a function of probe power. Data with the states $|0\rangle$, $|1\rangle$ and $|2\rangle$ are demonstrated respectively, while blue dashed line indicates the working point.  }
	\label{s2}	
\end{figure}

	The sample used in our experiments is a 3D transmon, as shown in Fig. \ref{s2},
		which consists of a superconducting qubit embedded in a 3D aluminium cavity
		\cite{paik_3d}. The cavity in our experiments is used to
		provide a convenient method to manipulate and measure the qubit. We employed
		an experimental setup for manipulating and measuring of the 3D
		transmon. Basically, there are two SMA connectors on
		the 3D cavity for microwave input and output, respectively. The input
		(output) quality factor is adjusted to be about $5\times10^{5}$ ($2\times10^{4}).$
		Microwave pulses for manipulating and reading out qubit are sent in through
		input connector after appropriate attenuation and isolation. A microwave
		generator combined with an in-phase and quadrature (IQ) mixer can produce
		microwave pulses for qubit manipulating. By adjusting the voltage of the IQ
		mixer we can control the phase (i.e. X and Y components) of microwave. To
		read out qubit states, we use ordinary microwave heterodyne setup.
		The output microwave is pre-amplified by HEMT at 4 K
		stage in the dilution refrigerator and further amplified by two low-noise
		amplifiers at room temperature. The microwave is then tuned into 50 MHz and
		collected by ADCs. In order to simplify our experimental procedures and data
		analysis while maintaining sufficient signal-to-noise ratio, we choose
		\textquotedblleft high power readout" scheme \cite{Reed_readout}.
		 Simply speaking, we send in a strong microwave
		on-resonance with the cavity, and the transmitted amplitude of the microwave
		will reflect the state of qubit due to the non-linearity of the cavity QED
		system.
		
		 We first use saturation spectroscopies to determine the transmon parameters.
		The resonant peaks indicate that the transition frequencies between $%
		|i\rangle $ to $|j\rangle $ is $\omega _{01}/2\pi =$ 7.1194 GHz, $\omega
		_{12}/2\pi =$ 6.7747 GHz, and $\omega _{23}/2\pi =$ 6.3926 GHz. From these
		we obtain the Josephson energy $E_{J}/\hbar \sim $ $2\pi \times 20.71$ GHz
		and the charge energy $E_{C}/\hbar \sim 2\pi \times 0.341$ GHz. The bare
		resonant frequency of the cavity is 9.0526 GHz. We measure the energy
		relaxation times of the energy levels $|1\rangle ,$
        $|2\rangle ,$ and $|3\rangle $ using the pump-and-decay method. It is found
		that $T_{1}^{01}\sim 15$ $\mu s$ , $T_{1}^{12}\sim 12$ $\mu s,$ and $%
		T_{1}^{23}\sim 10$ $\mu s$, respectively. The dephasing times are obtained from
		Ramsey measurement, which are $T^{*01}_2\sim$6.0 $\mu s$, $T^{*01}_2\sim$4.5 $\mu s$ and $T^{*01}_2\sim$3.1 $\mu s$, respectively.
	
In the experiments, we have to accurately design the magnitude, frequency and phase of the microwaves, which can be controlled by the waveform pulses applied to IQ mixer. We also calibrate the amplitude, phase and offset of the applied pulses. By adjusting these parameters carefully, we can significantly suppress the leakage of the LO signal and sideband mirror.  Tomography results indicate that the performance of our IQ mixer is very good.

	\subsection{The system Hamiltonian}

	The quantum  system  which consists of a transmon dispersively coupled to a cavity can be described by the Jaynes-Cummings model, the Hamiltonian can be written as:
	\begin{equation}
	H=\hbar\omega_{r} a^\dagger a+\sum_{i} \hbar \omega_{i}|i\rangle\langle i|+ \sum_{i,j} (\hbar g_{i,j}a^\dagger|i\rangle\langle j|+h.c.),
	\label{hamJC}
	\end{equation}
	where $\omega_{r}$ is the frequency of bare cavity, and $a$ $(a^\dagger)$ is the annihilation (creation) operator of the photon field. The transmon transition frequency $\omega_{i,i+1}=\omega_{i+1}-\omega_{i}$ is largely detuned from cavity frequency, and $g_{i,j}$ denotes coupling strength to the transition  between $|i\rangle$ and $|j\rangle$. In the dispersive limit, where $g_{i,i+1} \ll \omega_{r}-\omega_{i,j}$, Eq. (\ref{hamJC}) can be transformed to
	\begin{equation}	
    H=\hbar\omega_{r} a^\dagger a+\sum_{i=0}^{n} \hbar \omega_{i}|i\rangle\langle i|+ \sum_{i=1}^{n-1} \hbar \chi_{i-1,i}|i\rangle\langle i|+\sum_{i=1}^{n-1}\hbar(\chi_{i-1,i}-\chi_{i,i+1})|i\rangle\langle i|a^\dagger a,
    \label{dressed}
    \end{equation}
	where  $\chi_{ij}=g^2_{ij}/\Delta_{ij}$,  and $\Delta_{ij}=\omega_{ij}-\omega_{r}$ is detuning from the cavity frequency. Since the cavity is treated as the detector of transmon in the dispersively coupled quantum system, we can ignore the Hamiltonian of the cavity in Eq. (\ref{dressed}). In our experiments, if we apply two microwaves to couple the lowest three energy levels $\{|0\rangle,|1\rangle,|2\rangle\}$, the Hamiltonian in Eq. (\ref{dressed})  can be rewritten as ($\hbar=1$)
	\begin{equation}
	H_T=\omega_{0}|0\rangle\langle 0|+ \omega'_{1}|1\rangle\langle 1|+ \omega'_{2}|2\rangle\langle 2|+ (\Omega_{01}\cos(\omega_{m1}t+\phi_{01}) |0\rangle\langle 1|+h.c.)+(\Omega_{12}\cos(\omega_{m2}t+\phi_{12})|1\rangle\langle 2|+h.c.).
	 \end{equation}
	Here $\omega'_{i}=\omega_{i}+\chi_{i-1,i}+(\chi_{i-1,i}-\chi_{i,i+1})\hat{n}$ with $\hat{n}$ being the photon number in quantum cavity. $\Omega_{ij}$, $\omega_{mi}$ and $\phi_{ij}$ correspond to the amplitude, frequency and phase of the applied microwaves.
	In the interaction picture,  $H^I_T=\hat{U} H_T \hat{U}^\dagger-i(\partial_t \hat{U}^\dagger) \hat{U} $, where $\hat{U}=|0\rangle\langle 0|+\exp(i\omega'_{01}t)|1\rangle\langle 1|+\exp(i\omega'_{12}t+i\omega'_{01}t)|2\rangle\langle 2|$, $\omega'_{01}=\omega'_{1}-\omega_0$ and $\omega'_{12}=\omega'_{2}-\omega'_{1}$. Using the rotating-wave approximation, we can simplify the Hamiltonian as below

	\begin{equation}
	H^I_T=\delta_{1}|1\rangle\langle 1|+ (\delta_{1}+\delta_{2})|2\rangle\langle 2|+ \frac{1}{2}(\Omega_{01}e^{-i\phi_{01}} |0\rangle\langle 1|+h.c.)+ \frac{1}{2}(\Omega_{12}e^{-i\phi_{12}}|1\rangle\langle 2|+h.c.),
	\end{equation}
    where $\delta_{i}=\omega_{mi}-\omega_{i-1,i}(i=1,2)$. By carefully designing the parameters of the applied microwaves, we can map $H^I_T$ to desire the target Hamiltonian. For instance, with $\delta_{i}=0$, $\Omega_{01}\cos(\phi_{01})=\Omega q_x$, $\Omega_{01}\sin(\phi_{01})=\Omega q_y$, $\Omega_{12}\cos(\phi_{12})=\Omega q_z$ and $\Omega_{12}\sin(\phi_{12})=-\Omega q_w$ ($\Omega$ is the energy unit), we can construct the Weyl-like Hamiltonian in the main text,
	\begin{eqnarray}
	\begin{aligned}
	H_{4D}=\frac{\Omega}{2}\begin{bmatrix}
	0&q_x - i q_y&0 \\
	q_x + i q_y&0&q_z + i q_w \\
	0&q_z - i q_w&0
	\end{bmatrix}.
	\end{aligned}
	\end{eqnarray}

	\subsection{Energy structure measurement in a multi-level system}
	
	\begin{figure}[tbph]
		\includegraphics[width=14cm]{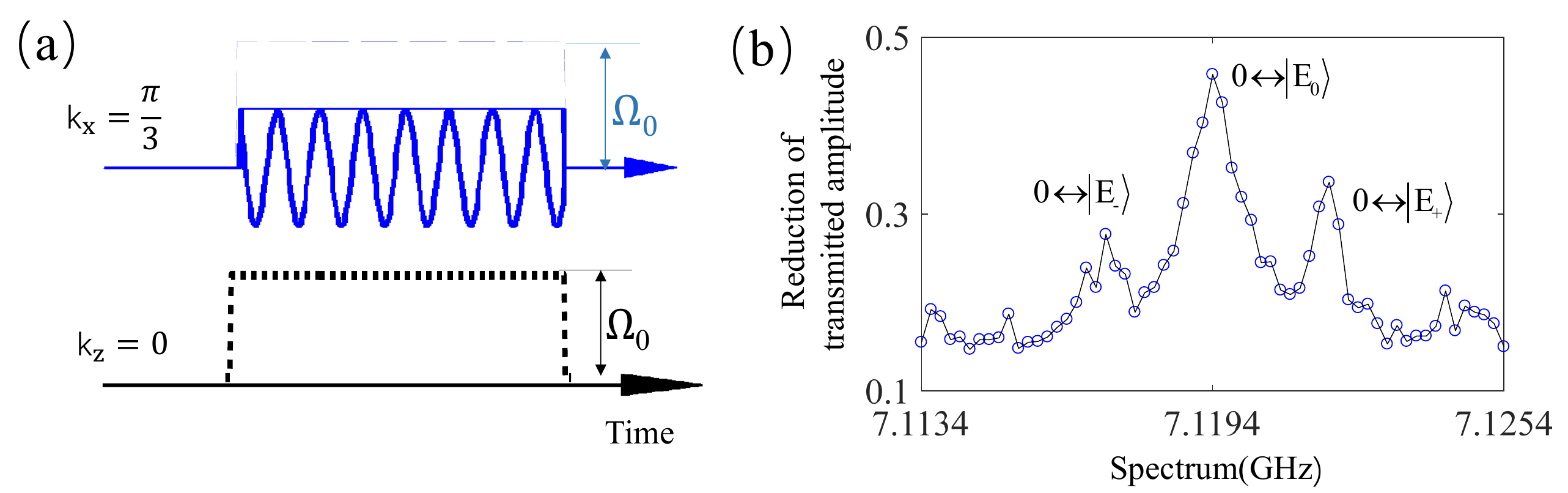}
		\caption{(a) The waveform of construct microwave send to the
			transmon for the typical parameters $k_{x}=\protect\pi /3$ and $k_{z}\approx0$. The
			blue (black) pulse corresponds the $\Omega^{1}_{x}$ ($\Omega^{2}_{x}$) of driving microwave.
			(b) An example of the spectrum of fixed $k_{x}$ and $k_{z}.$ The
			driving of construct microwave transforms the bare states ($|1\rangle $, $%
			|2\rangle $, and $|3\rangle $) to the eigen-states of driven system (i.e.,
			dressed states) ($|\psi_0\rangle $, $|\psi_+\rangle $, and $|\psi_-\rangle )$. By
			sweeping the frequency of the probe microwave (horizontal axis), we can observe
			resonant peaks at frequencies corresponding eigen-energies of dressed
			states, from which the energy structure can be extracted.
			As described in the main text, we only measure populations of the dressed states $|\psi_0\rangle $ and $|\psi_\pm\rangle $ distributed at the bare state $|1\rangle $.
	      }
		\label{spectrum}
	\end{figure}

When measuring the energy spectra in our experiments, we turn to use $|1\rangle $, $|2\rangle $ and $|3\rangle $ of transmon to form an artificial spin-1 particle and $|0\rangle$ is treated as a reference level for spectrum probing, as shown in Fig. 2(a) in the main text. First of all, the whole system is initialized in the ground state $|0\rangle $. The microwaves with frequencies $\omega_{12} $ and $\omega_{23}$ are then applied
to generate the transitions $\Omega^{1}_x$, $\Omega^{1}_y$, $\Omega^{2}_x$ and $\Omega^{2}_y$ and construct the Hamiltonian $H_{\text{exp}}$:

\begin{eqnarray}
\begin{aligned}
H_{\text{exp}}
=\frac{1}{2}\begin{bmatrix}
0&\Omega^{1}_{x}-i \Omega^{1}_{y}&0 \\
\Omega^{1}_{x}+i \Omega^{1}_{y}&0&\Omega^{2}_{x}+i \Omega^{2}_{y} \\
0&\Omega^{2}_{x}-i \Omega^{2}_{y}&0
\end{bmatrix},
\end{aligned}
\label{Ham_exp1}
\end{eqnarray}
where $\Omega^{1(2)}_{x}$ $(\Omega^{1(2)}_{y})$ is the Rabi frequency along the $x$ ($y$) axis of the Bloch sphere spanned by the corresponding basis. Since the four Rabi frequencies $\Omega^{1,2}_{x,y}$ are independently tunable, for simulating the Bloch Hamiltonian $H_k$ [see Eq. (\ref{Ham1})] in the 4D parameter space, they can be parameterized as
$\Omega^{1}_{x}=\Omega_0 (3+\Lambda -\cos k_x -\cos k_y -\cos k_z -\cos k_w)$, $\Omega^{1}_{y}=\Omega_0 \sin k_y$, $\Omega^{2}_{x}=\Omega_0 \sin k_z$, $\Omega^{2}_{y}=\Omega_0 \sin k_w$, where $\Omega_0=\sqrt{(\Omega^{1}_x)^{2}+(\Omega^{1}_y)^{2}+(\Omega^{2}_x)^{2}+(\Omega^{2}_y)^{2}}$.

The corresponding energy levels of the Hamiltonian in Eq. (\ref{Ham_exp1}) are obtained by measuring the eigenenergies of the
Hamiltonian. Measuring eigen-energies of the microwave-driven three-level
system is similar to that of the spectroscopy measurement with saturation
microwave, which is widely used in qubit experiments \cite{Tan2018}. The driven three-level
system forms dressed states $%
|\psi_0\rangle $ and $|\psi_\pm \rangle $, which can be written as
\begin{equation}
|\psi_+ \rangle=\frac{1}{\sqrt{2}} \begin{pmatrix} (\Omega^{2}_x+i\Omega^{2}_y)/{\Omega_0}
\\1
\\(\Omega^{1}_x+i\Omega^{1}_y)/{\Omega_0}
\end{pmatrix},
|\psi_0 \rangle=\begin{pmatrix}-(\Omega^{1}_x+i\Omega^{1}_y)/{\Omega_0}
\\0
\\(\Omega^{2}_x+i\Omega^{2}_y)/{\Omega_0}\end{pmatrix},
|\psi_- \rangle=\frac{1}{\sqrt{2}}
 \begin{pmatrix} (\Omega^{2}_x+i\Omega^{2}_y)/{\Omega_0}
\\-1
\\(\Omega^{1}_x+i\Omega^{1}_y)/{\Omega_0}\end{pmatrix}.
\label{Wfunction}
\end{equation}
The corresponding eigen-energies are $E_0=\omega_{01}$, $E_+ $=$\omega_{01}+\Omega/2$, and $E_-=\omega_{01}-\Omega/2 $, respectively. Figure \ref{spectrum}(a) is the example of selected
amplitude of microwave. Then we turn on the probe microwave. The widths of
the construct and probe microwaves are 200 $\mu s$ and 100 $\mu s,$ which
are much longer than the decoherence time of our transmon. We sweep the
frequency of probe microwave. When the probe frequency matches the energy
difference between an eigenstate and $|0\rangle ,$ the system will be
excited to the corresponding eigenstate. After turning off the construct and
probing microwaves, we sent a readout microwave pulse to the cavity to measure
the states of the system. Resonant peaks with frequencies representing the
eigen-energies have been observed as shown in Fig. \ref{spectrum}(b). Positions of resonant
peaks indicate the values of eigen-energies, while heights of resonant
peaks reflect each components of eigen-states at $|1\rangle$.
Then we change the parameter $k_{x}$ to collect the spectrum with
different resonant peaks. As shown in Fig. 2 in the main text, we shift the energy zero point to $\omega_{01}$.

We here discuss the spectral brightness distribution. In our experiments, the spectra we have measured actually reflect the populations of the dressed states $|\psi_0\rangle $, $|\psi_\pm\rangle $ at the bare state $|1\rangle$. The distributions of the dressed states at the bare states $|2\rangle$ and $|3\rangle$
are not necessary to measure, and furthermore they are more complicated to be measured since the two/three photon procedure is included. From Eq. (\ref{Wfunction}), we know that the brightness should be proportional to $P_\pm=|\langle1|\psi_\pm\rangle|^2=((\Omega^{2}_x)^{2}+(\Omega^{2}_y)^{2})/2\Omega_0^{2}$ and $P_0=|\langle1|\psi_0\rangle|^2=((\Omega^{1}_x)^{2}+(\Omega^{1}_y)^{2})/\Omega_0^{2}$.
Therefore, the ratio of the brightness are given by $P_+/P_0=P_-/P_0=(\Omega^{2}_x)^{2}+(\Omega^{2}_y)^{2})/2((\Omega^{1}_x)^{2}+(\Omega^{1}_y)^{2})$.

\begin{figure}[tbph]
	\includegraphics[width=10cm]{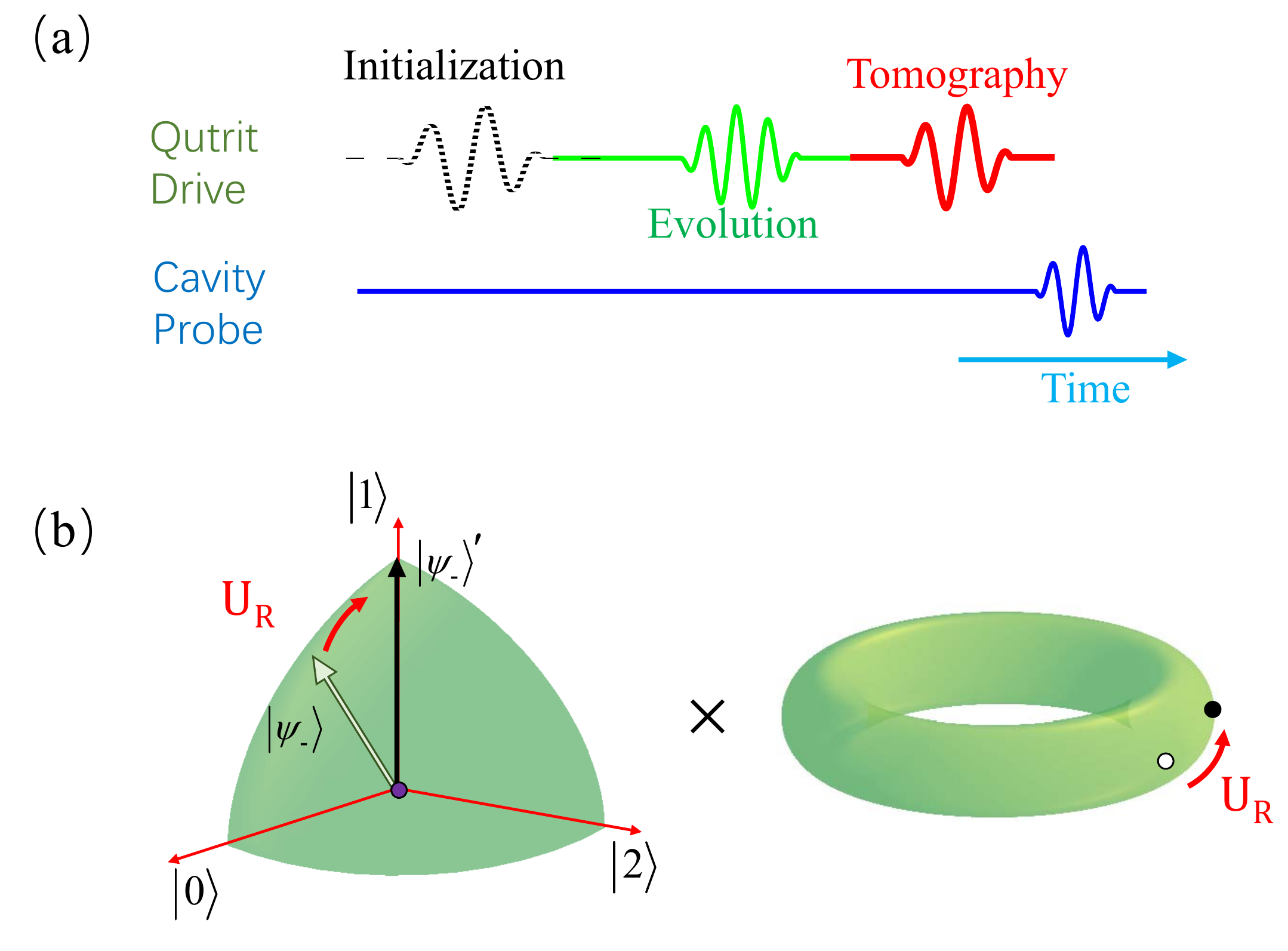}
	\caption{Scheme of measuring the quantum metric with sudden quench. (a) Schematic of the experimental procedures. Cavity signal is probed after pulse sequence of qutrit drive. (b) Illustration of $U_R$ rotation in Hilbert space of qutrit system, which is composed by an octant of a unit sphere and a torus \cite{Qutrit_Du,Qutrit_Wallraff}. Rotation $U_R$ of frame axis is applied in experiments to ensure the initial Hamiltonian $H'_0$ lying along the $\lambda_3$ axis. }
	\label{MetricFig}
\end{figure}

\subsection{Measurement of quantum metric using sudden quench}

To measure the probability of excited states, we have to perform quantum tomography of the qutrit as shown in Fig. \ref{MetricFig}, which is realized by measuring the density matrix $\rho$. Reconstruction of the full density matrix needs to do a set of rotations, $\textbf{I}$, $(\pm\frac{\pi}{2})^{01}_x$, $(\pm\frac{\pi}{2})^{01}_y$, $\pm\pi^{01}_x$,$\pm\pi^{01}_y$, $(\pm\frac{\pi}{2})^{12}_x$, $(\pm\frac{\pi}{2})^{12}_y$, $(\pi)^{01}_x(\pm\frac{\pi}{2})^{12}_x$, $(\pi)^{01}_x(\pm\frac{\pi}{2})^{12}_y$ and $(\pi)^{01}_x(\pm\pi)^{12}_y$. Here $I$ denotes the identical operation and $\theta_a^{ij}$ denotes a rotation along the axis with an angle $\theta$ on $ij$ transition, which contains more procedure than qubit tomography. After measuring the density matrix $\rho$, we calculate the probability of the eigen-state using

\begin{equation}
P=\langle \psi_- |\rho| \psi_- \rangle ,
\end{equation}
where $|\psi_-\rangle$ is the ground state of the driving Hamiltonian. To increase the accuracy of our experimental data, we repeat the measurements 16000 times and obtain the density matrix of the qudit  using the least square method. $|\psi_-\rangle$ is usually a superposition state of $|0\rangle$, $|1\rangle$ and $|2\rangle$, which is the function of $\theta_1$ and $\theta_2$. State preparation and tomography bring extra errors to practical 3-level experiments. To simplify state initialization and increase measurement fidelity, we rotate the frame axis with $U_R$ in the experiment to maintain the eigenstate of the initial Hamiltonian at the energy level $|0\rangle$ \cite{Tan2019b}, as shown in Fig.~\ref{MetricFig}. The extra benefit of this process is dramatic reduction of the decoherence effect. Consequently, the ramping Hamiltonian transforms to $H'(\boldsymbol q)=U_R H(\boldsymbol q) U_R^\dagger$. In practice, the $U_R$ can be decomposed as $U_R=\hat{R}(\beta_1)^z_{02}\hat{R}(\alpha_1)^y_{02}\hat{R}(\beta_2)^z_{01}\hat{R}(\alpha_2)^y_{01}$, where the operator $\hat{R}^{\hat{n}}_{ij}$ denotes a rotation along the axis $\hat{n}$ in the Bloch sphere spanned by the basis $\{|i\rangle, |j\rangle\}$. For an initial state $\frac{1}{\sqrt{2}}[\cos\theta_1-i\sin\theta_1\cos\theta_2,-1,\sin\theta_1\sin\theta_2]^T$, $\tan\alpha_1=-\sin\theta_1\sin\theta_2/|\cos\theta_1-i\sin\theta_1\cos\theta_2|,\tan\alpha_2=-(\sin\theta_1\sin\theta_2)^2-|\cos\theta_1-i\sin\theta_1\cos\theta_2|^2$, $\tan\beta_1=\sin\theta_1\cos\theta_2/\cos\theta_1$ and $\beta_2=-\beta_1/2$. If we write $H_0$ in the form as
	\begin{eqnarray}
	\begin{aligned}
	H_{0}
	=&\frac{1}{2}\begin{bmatrix}
	0&\Omega^{1}(t)e^{i\phi(t)}&0 \\
	\Omega^{1}(t)e^{-i\phi(t)}&0&\Omega^{2}(t) \\
	0&\Omega^{2}(t)&0
	\end{bmatrix},
	\end{aligned}
	\label{Ham_exp}
	\end{eqnarray}
	we can obtain the modified Hamiltonian
	\begin{eqnarray}
	\begin{aligned}
	H'_{0}
	=&\frac{1}{2}\begin{bmatrix}
(A^{\star}e^{2i\beta_2}+Ae^{-2i\beta_2})\sin\alpha_2\cos\alpha_2
	&-A^{\star}e^{2i\beta_2}\sin^2\alpha_2+Ae^{-2i\beta_2}\cos^2\alpha_2
	&-Be^{i\beta_2}\sin\alpha_2 \\
	A^{\star}e^{2i\beta_2}\cos^2\alpha_2-Ae^{-2i\beta_2}\sin^2\alpha_2
	&-(A^{\star}e^{2i\beta_2}+Ae^{-2i\beta_2})\sin\alpha_2\cos\alpha_2
	&Be^{i\beta_2}\cos\alpha_2 \\
	-Be^{-i\beta_2}\sin\alpha_2
	&Be^{-i\beta_2}\cos\alpha_2&0
	\end{bmatrix},
	\end{aligned}
	\label{Ham_exp}
	\end{eqnarray}
	where $A=\Omega^{1}(t)\cos\alpha_1e^{-i(\phi(t)-\beta_1)}-\Omega^{2}(t)\sin\alpha_1e^{-i\beta_1}$ and $B=\Omega^{1}(t)\sin\alpha_1e^{i(\phi(t)-\beta_1)}+\Omega^{2}(t)\cos\alpha_1e^{i\beta_1}$. For the Hamiltonian $H_{\text{exp}}$ with offset term $\Lambda$, we execute the same rotate procedure with modified parameters $\alpha_1(\Lambda)$, $\alpha_2(\Lambda)$, $\beta_1(\Lambda)$ and $\beta_2(\Lambda)$.

\end{document}